\documentclass[12pt, preprint]{aastex}
\usepackage{graphicx}
\usepackage{url}
\shorttitle{Diagnostics on the source properties of type II radio burst with spectral bumps}

\shortauthors{Feng et al.}

\usepackage{color}

\begin{document}

\title{Diagnostics on the source properties of type II radio burst with spectral bumps}

\author{S.W. Feng\altaffilmark{1,2,3}, Y. Chen\altaffilmark{1},
X.L. Kong\altaffilmark{1}, G. Li\altaffilmark{4,1}, H.Q. Song
\altaffilmark{1}, X. S. Feng\altaffilmark{2}, {\&} Fan
Guo\altaffilmark{5}} \altaffiltext{1}{Institute of Space Sciences
and School of Space Science and Physics, Shandong University,
Weihai 264209, China} \altaffiltext{2}{SIGMA
Weather Group, State Key laboratory for Space Weather, Center for
Space Science and Applied Research, Chinese Academy of Sciences,
Beijing 100190, China} \altaffiltext{3}{College of Earth Sciences,
Graduate School of Chinese Academy of Sciences, Beijing 100049,
China} \altaffiltext{4}{Department of Physics and CSPAR,
University of Alabama in Huntsville} \altaffiltext{5}{Department
of Planetary Sciences and Lunar and Planetary laboratory,
University of Arizona, Tucson, AZ 85721, USA}

\begin{abstract}
In recent studies (Feng et al., 2012; Kong et al., 2012),
 we proposed that source properties of type II
radio bursts can be inferred through a causal relationship between
the special shape of the type II dynamic spectrum (e.g., bump or
break) and simultaneous extreme ultraviolet (EUV)/white light imaging observations
(e.g., CME-shock crossing streamer structures). As a further
extension of these studies, in this paper we examine the CME event
dated on December 31 2007 associated with a multiple type
II radio burst. We identify the presence of two spectral bump
features on the observed dynamic spectrum. By combining
observational analyses of the radio spectral observations and the
EUV-white light imaging data, we conclude that the two spectral
bumps are resulted from CME-shock propagating across dense
streamers on the southern and northern sides of the CME,
respectively. It is inferred that the corresponding two type II emissions
originate separately from the two CME-shock flanks where the
shock geometries are likely quasi-perpendicular or oblique. Since
the emission lanes are bumped as a whole within a relatively short
time, it suggests that the type IIs with bumps of the study are
emitted from spatially confined sources (with a projected lateral
dimension smaller than 0.05-0.1 R$_\odot$ at a fundamental frequency
 level of 20-30 MHz).

\end{abstract}


\section{Introduction}

It is generally believed that metric type II solar radio bursts
are excited by energetic electrons accelerated at coronal shocks
(e.g., Wild, 1950; Dulk, 1985; Pick \& Vilmer, 2008). Although
extensive studies exist in the literature investigating the
physics of type IIs, the exact emission site and the associated
shock properties remain controversial mainly due to a lack of
direct high-resolution imagings of the bursts. The focus of the
debate lies in whether type II shocks are driven by flare heating
or coronal mass ejection (CME), (e.g., Wagner {\&}
McQueen, 1983; Gosling, 1993; Gopalswamy et al., 1998; Cliver et
al., 1999, 2005; Mancuso {\&} Raymond, 2004; Cane
{\&} Erickson, 2005; Vr\v snak {\&} Cliver, 2008;
Magdaleni\' c et al., 2008; Pohjolainen et al., 2008a), and whether the radio
\textbf{bursts are generated at the shock nose or at the shock flank}  (e.g. Reiner et al.,
2003; Mancuso {\&} Raymond, 2004; Cho et al., 2007, 2008, 2011;
Shen et al., 2013).

Recently we have proposed that source properties including the
emission site as well as the shock geometry can be inferred by
combining radio spectral data and solar imaging observations (Feng
et al., 2012; Kong et al., 2012). The main idea is to establish the physical
connection between certain morphological features of the type II spectrum (e.g., bump or
break) and the specific eruptive processes as recorded by coronal
imaging instruments (e.g., shocks crossing dense streamers).
Theoretical basis of this idea stems from the plasma emission hypothesis of type II
generation at coronal shocks, in which scenario the emission
frequencies and therefore the spectral shape is decided by the densities of
coronal structures along the shock path (Ginzburg \& Zheleznyakov,
1958).

Streamers are the most prominent bright large-scale structures in
the corona. They are expected to have effects on the type II radio
spectrum. Feng et al. (2012) and Kong et al. (2012) investigated
two CME-streamer interaction events with accompanying type II
radio bursts which occurred on November 1 2003 and March 27 2011. They
defined two special type II morphological features: spectral break
and spectral bump. According to their studies, a spectral break is
the result of density decrease when a type II emission
source propagates from inside of a streamer to outside,
and a spectral bump is produced when a type II source
propagates across a streamer structure from one
side to the other. Based on these  analyses, Feng et al.
(2012) and Kong et al. (2012) remarked that the emission site can
be pinpointed. Note that in some recent studies it has
been pointed out that when a shock crosses other localized
coronal and solar wind structures, such as dense coronal loops
(e.g., Pohjolainen et al., 2008b), pre-existing dense CME materials
and corotating interaction regions
(e.g., Knock {\&} Cairns, 2005; Schmidt {\&} Cairns, 2012;
Hillan et al., 2012), it may result in similar spectral shape changes.

To apply the above type II radio source diagnostic method to more
events, in this paper we examine another CME event which occurred
on December 31 2007. This event is associated with a multiple
type II radio burst (e.g., Robinson \&
Sheridan 1982; Shanmugaraju et al., 2005), well observed by the
imaging instruments on board
the Solar TErrestrial RElations Observatory (STEREO) A and B
(SA and SB for short) spacecraft (Kaiser et al., 2008) as well as
the Solar and Heliospheric Observatory (SOHO) spacecraft. The
event was associated with a radio burst
containing clear and rich signals including type III and type IV
bursts, and several episodes of type IIs. Many authors have
investigated different aspects of this event (Autunes et al.,
2009; Gopalswamy et al., 2009; de Koning et al., 2009; Liu et al.,
2009a, 2009b; Odstrcil \& Pizzo, 2009; Dai et al., 2010; Lin et
al., 2010; Moran et al., 2010; Cho et al., 2011; Rigozo et al.,
2011). In particular, Liu et al. (2009a), Gopalswamy et al. (2009)
and Cho et al. (2011) have discussed the origin of the type II
bursts. Inspecting the radio dynamic spectrum of this event, we
recognize that spectral bumps, similar to that studied by Feng et
al. (2012), are present on the type II spectrum. Considering the
extensive interests in this event and the potential important role
of spectral bump in revealing the type II origin, in this paper we
re-examine this event. Different from all previous studies, we
focus on the type II spectral bump and discuss how this feature
can be used to shed new lights on the type II origin.

\section{General properties of the event and previous studies}

In Figures 1a-1c we present three sets of EUV and white light
observations of the CME from SA, SOHO, and SB. The angle between SA and SB was 44$^\circ$ at the
time of this event. The superposed Inner Coronagraph (COR1) at $\sim$
01:00 UT and Extreme UltraViolet Imager (EUVI) at $\sim$ 00:55 UT
images from SA and SB and the difference Large Angle and Spectrometric
Coronagraph (LASCO) image together with the  Extreme-ultraViolet Imaging
Telescope (EIT) data from SOHO are shown. Black arrows denote the
eruption source active region (AR)10980, which is located at about E102S08,
E58S08, and E81S08 as viewed from SA, SB, and SOHO
(\url{http://www.lmsal.com/solarsoft/}). So, the CME is
seen  as a near-limb event by all three spacecraft. This
greatly constrains the projection effect on the measurements of
the CME dynamics. The CME is first observed by COR1A and COR1B at
00:55 UT and by LASCO C2 at 01:31 UT. The corresponding CME
leading edges are located at 1.55, 1.65, and 4.80 R$_\odot$,
respectively. A C8.3 flare is associated with the CME. According
to the GOES observation, the X-ray flux starts to increase
rapidly at 00:30 UT, and reaches its peak at 00:50 UT.

As clearly seen from the SA and SOHO observations,
streamers are present on both sides of the CME source.
 Figures 1d and 1e present the coronal magnetic field
configurations from the potential-field source-surface model
(PFSS; Schatten et al., 1969; Schrijver \& Derosa 2003) based on
the magnetic field measurements with Michelson Doppler
Imager (MDI; Scherrer et al., 1995) for the Carrington Rotation
2065. The magnetic configurations have been rotated to the view
angles of SA and SB, respectively. Large-scale closed field lines,
corresponding to the white light streamers, can be seen on both
sides of the active regions. The northern streamer is narrower and
weaker than the southern one. This is consistent with the fact
that the northern streamer is a pseudo streamer (PS: Wang et al., 2007;
 Liu et al., 2009b) while the southern streamer is a typical helmet streamer (HS).
For a PS, the polarities of the open field enclosing
the streamer-like structure are the same, as seen from the PFSS
results. In Figure 1a, we delineate the
estimated overall magnetic topologies of both streamers with cyan
lines.

Several additional features in Figures 1a-1c worth to be noted.
First, as seen from the LASCO difference image, there is a diffuse
structure ahead of the bright CME front. Such a structure
is usually regarded  as the signature of shock sheath (Vourlidas
et al., 2003). Small white arrows point to different locations of
the sheath front. As a result of the asymmetric expansion of the
CME, the stand-off distance varies significantly along the front.
Second, the PS is deflected
significantly as shown from the obvious white-black difference structure,
even in the absence of direct contact with
the bright ejecta. Streamer deflection without a CME
contact has been attributed to the CME-driven shock (e.g., Sheeley
et al., 2000). Finally, there is a concave-outward structure at
the CME front along the direction of the stalk of the southern
streamer, which is usually interpreted as a result of a CME
propagating into denser and slower plasma sheet structure (see,
e.g., Riley {\&} Crokker, 2004; Odstrcil et al., 2004). This
concave-outward structure therefore indicates a very strong
interaction of the CME with this streamer.
As to be shown below, this observation is consistent with the SA
data.

Now we turn our attention to the dynamic spectrum of the
associated radio burst, which was recorded by Learmonth (Kennewell \&
Steward 2003) and BIRS (Bruny Island Radio Spectrometer; Erickson
1997) and shown in the left panel of Figure 2. An enlargement of
the spectrum in the square region is given in the right panel with
the y-axis on a linear scale. Several type II episodes are
observed. The first episode shows fundamental (F) and
harmonic (H) branches with clear signatures of band
splitting on the H branch. It starts at $\sim$ 00:53 UT and ends
at $\sim$ 01:20 UT lasting for about 30 minutes. The H branch
spans from $>$ 100 MHz to 14 MHz.

The other two episodes are denoted by ``a'' (01:04 UT -
01:10 UT, 85 MHz - 35 MHz) and ``b'' (01:11 UT - 01:14 UT, 57 MHz
- 40 MHz) in Figure 2. These two emissions were regarded as
the fundamental and harmonic branches of a type II burst by
Gopalswamy et al. (2009) and Cho et al. (2011). However, with a
careful examination we conclude that this may not be the case. Note
that there are no temporally-overlapping emissions between ``a'' and ``b''.
Therefore, to determine their frequency ratio we need to fit
the dynamic spectra to allow a
direct comparison of their frequencies. The fitting curve to ``a''
from 01:08 UT - 01:11 UT \textbf{is drawn by} the white solid line in
Figure 2, using the 0.4$\times$ Saito density model
(Saito et al., 1977). \textbf{The fitting gives a shock speed of
550 km s$^{-1}$ (the reason of employing such a density model is
explained in the Appendix). Two white dashed lines with frequencies
1.8 and 2 times larger than the fitted curve are also plotted.} It can be seen that the
frequency ratio between the emission denoted by ``b'' and the fitting to
``a'' is less than $1.8$.
This excludes the possibility of ``b'' being the harmonic
counterpart of ``a'' (Nelson \& Melrose, 1985; Mann et al., 1995, 1996).
Therefore, we conclude that this multiple
type II event consists of three separate episodes of emissions.
\textbf{Since at metric wavelengths the harmonic emission is usually
brighter than the fundamental one (e.g., Cairns \& Robinson, 1987), we
regard ``a'' to be a harmonic-band emission.}


At $\sim$ 01:00 UT the lower band of the H branch of the first episode
($\sim$ 48 MHz) become discontinuous. Shortly after this, the band rises up to
a higher frequency of $\sim$ 52 MHz at 01:02 UT, then its frequency
decreases rapidly to $\sim 37$ MHz at 01:05 UT. Such a
non-monotonic variation of frequencies has been referred to as type
II spectral bump by Feng et al. (2012). Another bumping feature,
similar in shape, but weaker in amplitude, is present on ``a''
spanning from $\sim$01:06:10 UT - $\sim$01:07:40 UT around 60 MHz. The two
bumps are indicated by white arrows in the figure. Their origin
and physical implication are the focus of this study.

Before further discussion, we summarize relevant results from
previous studies on this type II event by Liu et al. (2009a),
Gopalswamy et al. (2009), and Cho et al. (2011).

Liu et al. (2009a) focused on the driver of the first type II
episode. They used the 1.3$\times$ Saito density model (Saito et al., 1977) with a shock
speed of 616 km s$^{-1}$ to fit the dynamic spectrum and deduce
the shock heights. The deduced heights were then compared to the
distance measurements of the CME front and the propagating streamer
kink induced by the CME shock-streamer interaction. This
established the physical connection between the metric and the
decametric-hectometric bursts. They concluded that this episode
was driven by the CME, rather than by the associated flare.
Gopalswamy et al. (2009) measured the height of the CME leading
edge at the time of the onset of the first episode and concluded
that the type II is emitted at a few tenths of a solar radius
above the solar surface. Cho et al. (2011) examined the type II
bursts as a multi-band event (e.g., Robinson \& Sheridan 1982; Shanmugaraju et al., 2005).
They used a Newkirk density model (Newkirk, 1961) to convert the
frequencies of the two episodes into shock heights. They found
that the obtained two sets of shock heights were consistent with
the measured CME-nose heights and the interaction heights of the
shock with the northern streamer, respectively. They then
concluded that the first episode was excited at the CME shock
nose, and the second episode ``a'' excited at the interaction
region of the shock with the northern streamer.

All of the above authors agreed that the type II burst was excited
by the CME-driven shock. None of them, however, discussed
the bumping features of the spectrum. As noted, these features
provide significant amount of information in diagnosing the origin
of the type II radio burst. In particular, one may infer the location
of the source and physical conditions for the generation of
type II radio bursts. We point out that features like these in
radio spectra should be carefully examined in relevant CME
studies. We note that the back-extrapolation of the second
episode ``a'' maps to the onset of a cluster of type III bursts, which
indicates a possible connection to the flare impulsive phase. Therefore,
a flare origin of this episode can not be ruled out completely. Nevertheless,
we only consider in this study the possibility that the type II emitting
shock is driven by the CME.

\section{CME-shock profiles and origin of the type II spectral
bumps}

In this section, we first introduce how we delineate the CME-shock
profiles from the EUVI and COR1 data so as to present a clear
picture of how the shock evolves and interacts with the streamers.
Then, we relate the deduced shock-streamer interactions to the
observed type II spectral bumps.

Coronal shock profiles can be determined directly using EUV and white light
imaging data. Observational signatures of a coronal shock include
diffuse sheath structure ahead of the bright CME ejecta (Vourlidas
et al., 2003), deflection and kink of streamer stalks and coronal
rays as swept by the CME shock (e.g., Sheeley et al., 2000), and
the EUV propagating disturbance revealing the expanding shock front.

In our event, both the diffuse sheath structure and streamer deflection/kink are
observed in the LASCO difference image at 01:32 UT. Since the streamers are
best seen by SA (see Figure 1), in Figure 3 we plot the eruption
sequence between 00:55 UT-01:15 UT as observed by COR1 and EUVI of
SA.  The upper and middle panels are the COR1 running difference and
direct images, respectively. The EUVI 195
{\AA} difference and direct data are included when available.
Corresponding animations can be viewed online.

In the upper panels of Figure 3, we use arrows of different color
to point out various shock signatures. The resultant continuous
shock profiles are plotted in the lower panels. The solid part of
the shock profiles is determined by recognizable shock signatures,
and the dotted part represents the profile without a clear shock
signature and is derived by assuming a smooth shock
propagation.

At 00:55 UT, a bright loop is observed by both EUVI
and COR1. This is the first appearance of the CME in the COR1
field of view. None of the above shock signatures are present at
this moment mostly because the shock is still rather close to the
bright ejecta at this very-early stage of the eruption. Note that
the shock is already formed since the corresponding
type II burst started earlier at 00:53 UT. We thus take the
position of the bright loop structure as the shock front location.
Five minutes later (01:00 UT), the loop propagates outward with a
fast lateral expansion. A diffuse sheath structure appears at the
northeastern quadrant as pointed by the upper arrows. No observable
sheath structure is present in the southeastern quadrant partially
due to the asymmetric propagation of the CME. The obtained shock
profile for 01:00 UT is shown in Figure 3g.

At 01:05 UT, the diffuse sheath structure is well observed ahead
of the ejecta in the northeastern quadrant. In addition, from the
EUVI 195 {\AA} data we can recognize the EUV fronts associated
with the eruption as indicated by the black arrows, which are
consistent with the deflection fronts of the EUV rays. We use these
features to determine the lower extension of the shock.
This allows us to plot a more complete shock envelope than at 01:00
UT. In the southeastern quadrant of the difference image, a weak
concave-outward structure along the direction of the stalk of the
southern streamer is discernible. The structure is more
prominent in the subsequent COR1 data. It was also seen
in LASCO, indicating a direct interaction between the CME and
the streamer, as mentioned.

At 01:10 UT and 01:15 UT, besides the diffuse sheath structure
(see white arrows) a new shock signature appears along the
northern streamer as indicated by the blue arrow. The streamer is
strongly deflected without a direct contact with the bright
ejecta, indicating a shock front propagating along this direction
(e.g., Sheeley et al., 2000).

At each time step, we connect the front of the diffuse structure,
the front of the streamer deflection, and the concave
outward-structure to obtain the full envelope of the CME-driven
shock (see Figures 3i and 3j). The obtained shock envelopes are
shown all-together in Figure 3k, where the circle represents the
solar limb on which the projected eruption source is plotted as
the plus sign. The magnetic configurations of the two streamers
are copied from Figure 1a. Also included are the shock profiles as
determined from the COR1A running difference images at 01:20 UT
and 01:25 UT, as well as that given by LASCO C2 at 01:32 UT. Since
the angular separation between SOHO and SA is about 22 degrees,
the shock profiles seen from SOHO and SA should be similar.
\textbf{This allows us to estimate} the shock speeds along
different directions. For example, the average speeds in the time
range of 00:55 UT - 01:10 UT along four directions pointing from the
eruption source \textbf{ (see arrows in Figure 3k) are estimated to 560, 620,
760, and 1000 km s$^{-1}$, respectively. These values reveal} a very asymmetric
propagation of the CME front. The blue dashed shock envelopes at
01:04 UT, 01:06:10 UT, and 01:07:40 UT are given by interpolations
between nearby shock profiles or extrapolations using the obtained
average shock speeds along relevant directions. Thus, Figure 3k presents a
relatively-complete description of the shock propagation from the
inner to the outer corona and provides important clues to
understanding the origin of type II spectral bumps.

Note that the presence of a spectral bump requires a high
density structure along the shock path according to the plasma
hypothesis of type II radio bursts (Ginzburg \& Zheleznyakov,
1958; Feng et al., 2012). From the coronagraph data alone, both the
northern and the southern streamers can be the candidate for the
two spectral bumps. From Figures 2 and 3 it can be seen that the
first bump ends at 01:04 UT, before the shock contacts the
northern PS ($\sim$01:05 UT). This rules out the possibility of
the PS accounting for this bump. \textbf{Further evidence supporting
the hypothesis that this spectral bump is caused by the shock transit
across the southern streamer can be found by examining the radio-source
height. Assuming a 1$\times$Saito density model (see the Appendix), we infer
that the radio-source at the onset ($\sim$48 MHz) and the end (37 MHz) of the
type II bursts was located at 1.69  R$_{\odot}$ and 1.83  R$_{\odot}$, respectively.}
 The two heights are presented in Figure 3k with two black dotted arcs. \textbf{The intersections
 of these arcs with the shock envelopes are indicated by small red ellipses},
 which are presumably the source locations of the corresponding type II episode at the onset and
the end of the bump.

\textbf{It can be seen that the two intersections are located at the opposite
sides of the pre-disturbed southern streamer.} This strongly indicates that the shock
transit across this streamer causes the first spectral bump.
In addition, from the bump duration ($ \tau \sim$ 4 minutes) and
the estimated shock speed in this region ($ v \sim$ 600 km s$^{-1}$,
see Figure 3k), one infers the width $ D \sim \tau v$ of the dense
structure to be 0.2 R$_\odot$, in agreement with the
white light data of the southern streamer. Based on these arguments,
we conclude that the first bump is caused by the shock transit
across the southern streamer.

The second bump starts at 01:06:10 UT and lasts for $\sim$1.5
minutes. Its beginning is coincident with the first interaction of the
shock with the northern PS. As mentioned earlier, the
corresponding type II emission is regarded as the harmonic branch. To
examine the role of the PS in the formation of the
spectral bump, we convert the frequencies at the beginning and end of
this bump to shock heights using the 0.4$\times$ Saito model
(which is appropriate to describe the density distribution near the
PS, see the Appendix). \textbf{The radio-source heights are found to
be 1.32 R$_\odot$ and 1.38 R$_\odot$ and we depict these height levels
by two dashed arcs in Figure 3k. Their intersections with the shock envelopes
at 1:06:10 UT and 1:07:40 UT, marked by small red ellipses, are found to be
located at the two sides of the pre-disturbed PS. This is similar to the situation
for the first bump. In addition, the product }of the bump duration ($\sim$1.5
minutes) and the estimated shock speed ($\sim$1000 km s$^{-1}$, see
Figure 3k) agrees with the observed streamer width, and the
fact that the PS is narrower in width and weaker in brightness (smaller in density) is also
consistent with the corresponding spectral bump being weaker in
amplitude and shorter in duration. We therefore conclude that the
shock transit across the PS causes the second bump.

In summary, by combining the dynamic spectral and imaging data we
are able to determine the physical origin of the
two spectral bumps. It suggests that the two type II episodes are
produced separately at the two flanks of the same CME shock. From
the shock envelopes and the estimated coronal magnetic field
topologies from the PFSS model, one finds that the shock geometry
at the two flanks are more perpendicular or oblique than
quasi-parallel.

Last, in Figure 2 we have presented the fitting curves
with the 1$\times$ and 0.4$\times$ Saito density models to the pre-bump
spectra of the first and the second episodes. From the
fittings we deduce the radio source speeds to be 530 km s$^{-1}$ for the first
episode and 650 km s$^{-1}$ for the second episode. Note
that in our event the \textbf{radio source propagates} along a highly non-radial direction,
 so these speeds are representative of the radial components.
These values agree with the speed measurements deduced
from the white light images as shown in Figure 3k, therefore providing a
self-consistency check of the validity of the density model used
for the fittings.

\section{Discussion on the inferred source size of type II bursts}

We point out that observations of spectral bumps of type II radio burst
can be employed as a unique method to infer the
source size of the type II radio bursts. From the dynamic spectrum
we see that the type II emission lane is raised up as a whole
within a relatively short interval. This implies that the size of
the radio source is smaller than the associated dense streamer
structure ($\sim$ 0.1 - 0.3 R$_\odot$). The short duration of the rising
part also suggests that the source is compact in
spatial extension. Due to the intermittency of the radio signals
at the onset of the first type II bump, we are unable to determine
the exact time duration for the spectral elevation. However, one
can infer an upper limit of $\sim 1$ minute. Then assuming the
shock crosses the streamer with a speed of $\sim$ 600-1000 km
s$^{-1}$, we deduce that the spatial dimension of the radio source
needs to be smaller than 0.05-0.1 R$_\odot$  at a fundamental frequency
level of 20-30 MHz. Comparing to the very broad extension
of the shock surface (easily $>$ 1 R$_\odot$), the
type II source can be practically regarded as ``point-like''.

Earlier data analyses of radioheliographs revealed that type II
sources were restricted to discrete sectors of an arc around the
flare center (Wild 1970; Wild \& Smerd, 1972), and sometimes of
large dimension ($\sim$ 0.5 R$_\odot$) (Kai \& McLean 1968).
It is well known that the radio source size obtained by radio-heliographs tends to
be larger than the real value due to propagation and scattering of
radiation from the source to the observer. Also, earlier radio
imaging instruments such as the Culgoora and Clark Lake
Radio-heliographs have an angular resolution of several
arcminutes at frequencies $<$ 100 MHz, which corresponds to a
spatial resolution of a few tenths of a solar radius near the Sun.
Therefore these instruments were not able to resolve the type II
source as small as that deduced from our study, although the
discrete feature of the imaging observations agrees with our
result.

The suggestion that the type II source is compact implies that
the presence of shocks is only a necessary condition for the
generation of type II bursts. Other strict physical conditions
must also be satisfied at the shock in order to create a
non-thermal distribution of electrons which is unstable to plasma
instabilities. For example, the radio-emitting shock front may be
quasi-perpendicular, as revealed by our study. Early theoretical
studies have proposed that quasi-perpendicular shocks can
accelerate electrons by the shock drift acceleration mechanism
(Holman \& Pesses 1983, Wu 1984). When the shock speed is
sufficiently large or $\theta_{Bn}$ (the angle between shock
normal and upstream magnetic field) is close to 90 degrees, some
electrons can be accelerated to non-thermal energy and excite
plasma waves. However it is known that in this process both the
fraction and achievable energy of the accelerated particles are
limited (e.g., Ball \& Melrose 2001). Other effects, such as MHD
turbulence, shock ripples and/or magnetic
collapsing trap geometries (e.g., Decker, 1990; Zlobec et al., 1993;
Magdaleni\'{c} et al., 2002; Guo {\&} Giacalone, 2010; Schmidt {\&}
Carins, 2012; Hillan et al., 2012) may be required to yield more
efficient electron acceleration and radio emission. If
indeed local structures, such as ripples or magnetic traps, play
an important role in accelerating electrons, then it is understandable
that the radio-emitting region at the shock front is spatially confined.

In space weather studies, the frequency drift of type II burst is
often employed as an important input to predict the shock arrival
time at Earth. However, if the type II burst is generated from a
special discrete part of the shock flank, as inferred from our
study, then the speeds obtained from the curve fitting to the
dynamic spectrum may not be accurate. This should be taken into
consideration when using type II spectrum as inputs to drive space
weather forecastings.

\section{Conclusions}

In this paper we investigated the physical origin of two bump
features on the dynamic spectrum of a multiple type II radio
burst, which were associated with a CME event occurred on December
31 2007. Combining radio spectral data and EUV/white light imaging
observations, we found that the type II bumps are caused by the
source density variation when the CME shock propagates through
nearby dense streamers. It is suggested that the two type II
episodes are generated separately at the two flanks of the
CME-driven shock. It is further inferred that the type II signals
are emitted from discrete spatially-confined sources at the CME-shock
flank with the source spatial extension smaller than 0.05-0.1 R$_\odot$
at a fundamental frequency level of 20-30 MHz
and the large-scale shock geometry is close to quasi-perpendicular and/or
 oblique.

%
%

\appendix

\section{Appendix: Coronal electron density distribution deduced with the pB inversion method}

The coronal electron density ($n_e$) distribution can be
deduced by inverting the \textbf{polarization brightness} (pB) data recorded
by coronagraphs. In Figure 4a, we show the pB data observed by
LASCO C2 at 21:05 UT on December 30 2007, $\sim$ 4 hours before the
\textbf{type II radio burst}. We assume that the coronal
background density distribution does not change significantly
during this period. Note that the COR1/2 coronagraphs on board
STEREO also record the pB data during this event, however, the
subtraction of background emission which is dominated by the
scattering of dusts on the objective lens does not allow one to
determine $n_e$ outside of dense streamer regions (c.f., Frazin et
al., 2012). Considering the separation angle between SOHO and
STEREO A is relatively small ($\sim$22 degrees) and the similarity
between the LASCO image and the COR1A image, we use the LASCO pB
data to derive $n_e$ for our study.

The standard pB inversion method given in the SolarSoft
package is used. Radial profiles of $n_e$ along three position
angles covering most of the CME expansion region are deduced and
plotted in Figure 4b. It can be seen that $n_e$ distributes
asymmetrically. \textbf{At the regions close to the southern and
northern streamer, $n_e$ can be well represented by 1$\times$ and
0.4$\times$ Saito density model (Saito et al., 1977), respectively.
Furthermore, $n_e$ generally decreases from}
the southern to the northern streamer, as indicated by
latitudinal variation of pB values at two distances (2.5 R$_\odot$ and
3 R$_\odot$ see Figure 4c). The projected angular width of
the southern (northern) streamer is $\sim$10 (5) degrees.

\acknowledgements We are grateful to the STEREO, SOHO/LASCO,
MLSO/MK4, Wind/Waves, BIRS, and LEAR teams for making their data
available to us. This work was supported by grants NSBRSF
2012CB825601, NNSFC 41274175, 40825014, 40890162, 41028004, and
the Specialized Research Fund for State Key Laboratory of Space
Weather in China. H.Q. Song by NNSFC 41104113 and 41274177. Gang Li's
work at UAHuntsivlle was supported by NSF grants ATM-0847719 and
AGS1135432. We thank the anomalous referee for constructive comments and suggestions.

\newpage
\begin{figure}
 \includegraphics[width=0.8\textwidth]{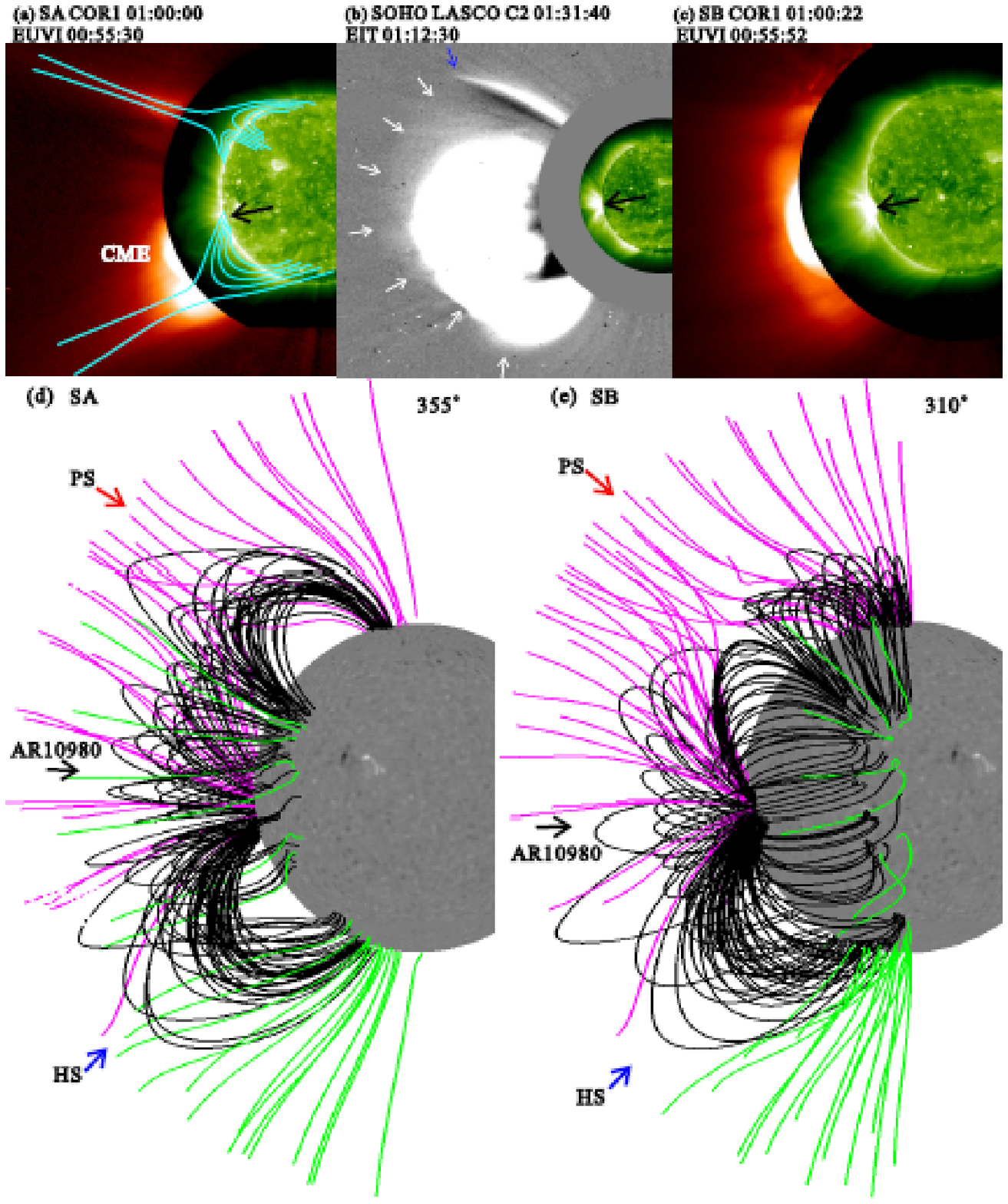}
\caption{(a-c) EUV and white light observations of the CME from
SA, SOHO, and SB on December 31 2007. The angle between SA and SB
was 44$^\circ$ at the time of this event. The COR1 ($\sim$
01:00 UT) and EUVI ($\sim$ 00:55 UT) images from SA and SB and the
difference LASCO image together with the EIT data from SOHO are
shown. Black arrows denote the eruption source AR10980, white
arrows denote the diffuse sheath structure ahead of the bright
ejecta, and the blue arrow points to the front of the streamer
deflection caused by the shock. (d-e)
 Coronal magnetic field configurations as given by the PFSS model
 based on the magnetic field measurements with MDI
 for the Carrington Rotation 2065. The magnetic configurations have been
 rotated to the view angles of SA and SB, respectively. The estimated
 overall magnetic topologies of the PS and HS
 are plotted with cyan lines in (a).} \label{Fig:fig1}
\end{figure}

\begin{figure}
 \includegraphics[width=1.0\textwidth]{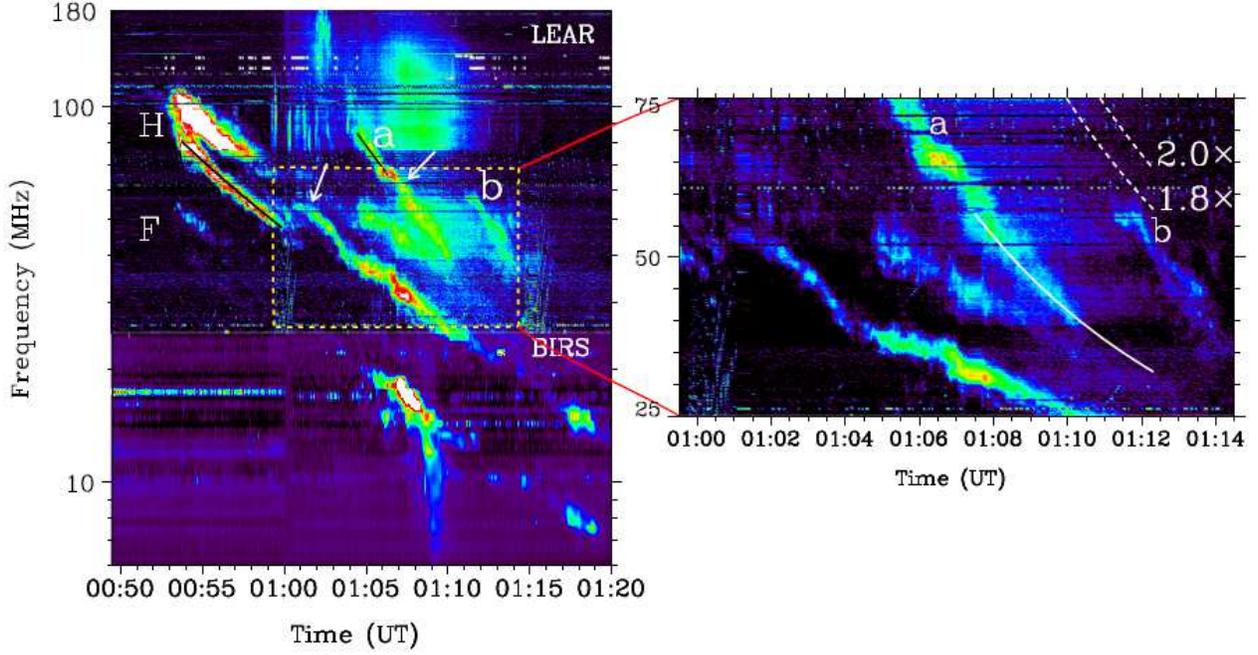}
\caption{Left: the dynamic spectrum of the associated multiple
type II  radio burst as recorded by Learmonth and BIRS radio spectrometers.
Three type II episodes are observed. The first episode shows fundamental
(F) and harmonic (H) branches with clear signatures of band splitting of H branch.
The other two episodes are denoted by ``a'' and ``b''. White arrows indicate the
spectral bump features on the first two episodes. The black
lines are the fitting curves using the 1$\times$ Saito density
model and a shock speed of 530 km s$^{-1}$ for the first episode and
650 km s$^{-1}$ for the second using 0.4$\times$ Saito density model.
 Right: an enlargement of the spectrum in square region of the left panel
 with the y-axis in a linear scale. The solid white line is the fitting curve
 of the post-bump part of ``a'' using 0.4$\times$ Saito density model and a shock
 speed of 550 km s$^{-1}$. The upper two dashed white lines are given by 1.8 and
 2 times larger than this fitting curve.} \label{Fig:fig2}
\end{figure}

\begin{figure}
 \includegraphics[width=0.7\textwidth]{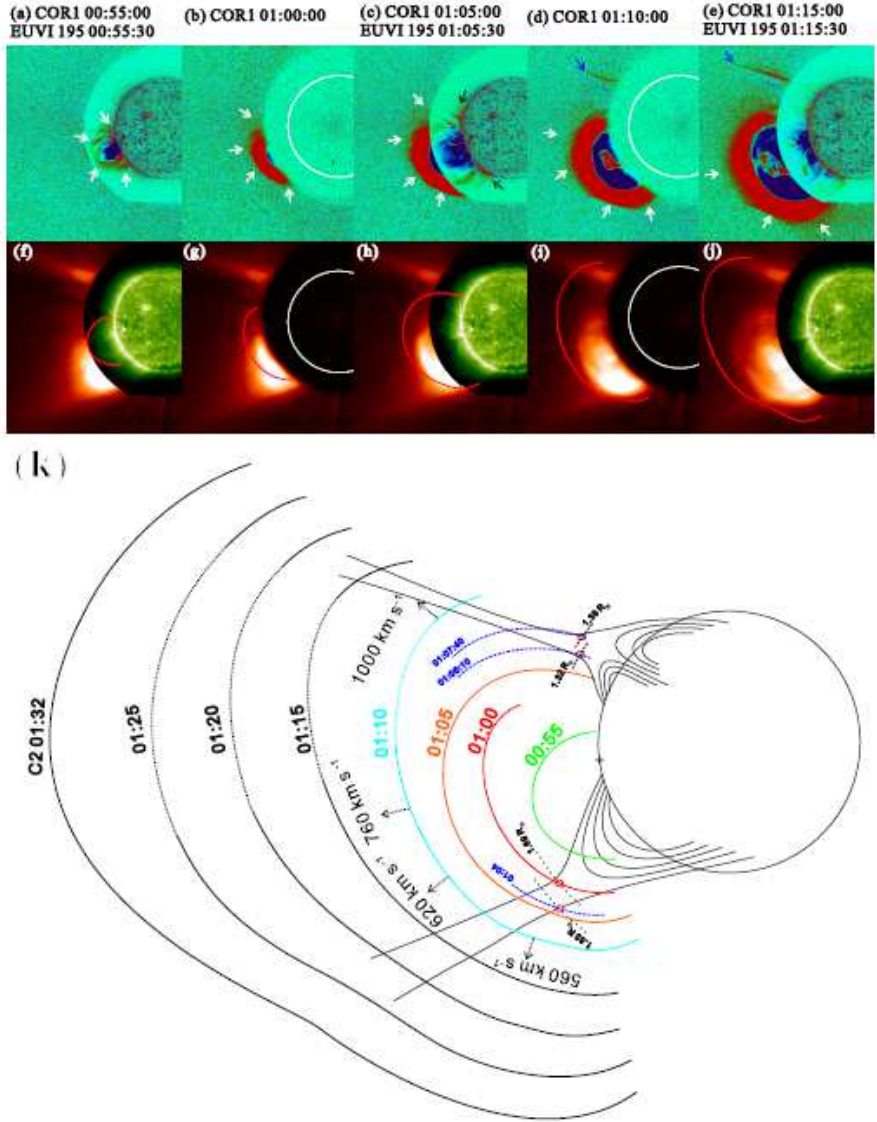}
\caption{(a-e, f-j) The eruption sequence between 00:55-01:15 UT
as observed by COR1 and EUVI of SA. The upper and middle panels
present the COR1 running difference images and direct images,
respectively, and the EUVI 195 {\AA}  difference and direct data are
included when available. In panels (a-e), we use arrows of
different colors to point out various shock signatures.
The resultant continuous shock profiles are plotted in panels (f-j).
 Corresponding animations can be viewed online. (k) The shock profiles
obtained from the above panels. The circle represents the solar limb.
The magnetic configurations of the northern and southern streamers are copied
from Figure 1a. Also included are the shock profiles as determined
from the COR1 running difference images at 01:20 UT and 01:25 UT
and the LASCO C2 data at 01:31 UT. The blue dashed shock profiles
at 01:04 UT, 01:06:10 UT, and 01:07:40 UT are given by interpolations
(or extrapolations) using the obtained average shock speeds along relevant directions
 of nearby shock profiles, and the two pairs of black dotted
arcs are given by r=1.69 R$_\odot$, 1.83 R$_\odot$
and r=1.32 R$_\odot$, 1.38 R$_\odot$. Small red ellipses represent the
intersection points of these arcs with the shock envelopes.
See text for more details.} \label{Fig:fig3}
\end{figure}

\begin{figure}
 \includegraphics[width=1.0\textwidth]{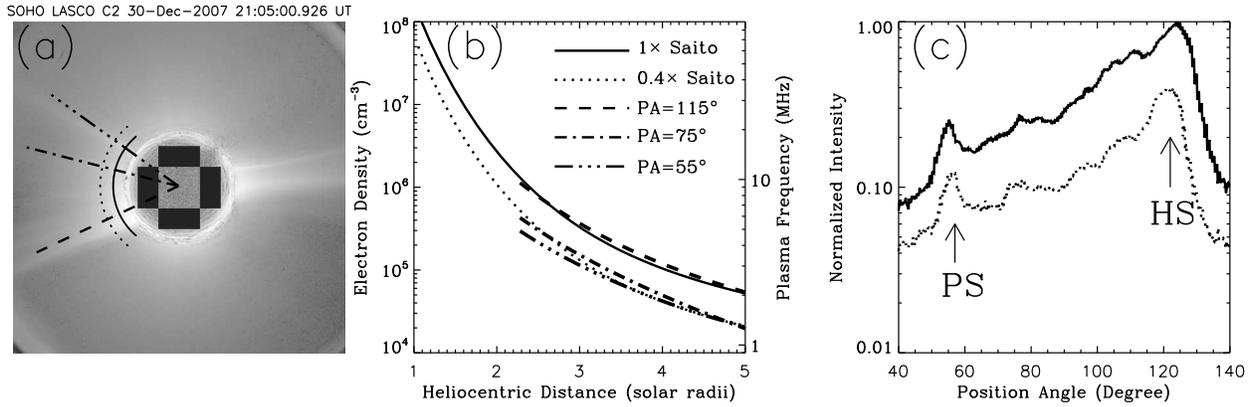}
\caption{(a) The pB data observed by LASCO C2 at 21:05 UT on December 30 2007;
(b) Radial profiles of $n_e$ along three position
angles as deduced with the standard pB inversion method,
\textbf{the density profiles given by 1$\times$ and 0.4$\times$ Saito density
model are depicted by the solid and dotted line, respectively};
(c) Latitudinal variations of normalized pB
intensity at two distances of 2.5 R$_\odot$ (solid) and 3.0 R$_\odot$ (dotted).}
\label{Fig:fig4}
\end{figure}

\end{document}